# The Upgrade of EAST Safety and Interlock System

Z.C, Zhang, B.J. Xiao, Z.S. Ji, Y. Wang, F. Xia and Z.H. Xu

*Abstract*—The Experimental Advanced Superconducting Tokamak (EAST), a nation-level large-scale scientific project of China, plays a key role for the research of peaceful utilizations of fusion energy. The safety and interlock system (SIS) is in charge of the supervision and control of all the EAST components involved in the protection of human and tokamak from potential accidents. With the development of physical experiment, the SIS had come close to reaching its limits for expandability. Therefore, a prototype for upgrading EAST SIS has been designed, and a fast architecture based on COTS FPGA is absorbed into the new SIS. This paper presents EAST machine and human protection mechanism and the architecture of the upgrading safety and interlock system.

## I. INTRODUCTION

NUCLEAR fusion which is a controllability reaction has unquestionably been recognized to be one of the best ways to solve the future energy crisis, and the tokamak is a favourable equipment to realize nuclear fusion [1]. The national project of experimental advanced superconducting tokamak (EAST) is an important part of the fusion development stratagem of China, which is the first fully superconducting tokamak with a non-circle cross-section of the vacuum vessel in the world. The huge tokamak project is complicated and high cost, needs to be monitored to eliminate the probability of potential hazard during the physical campaign [2]. The role of safety and interlock system (SIS) is focuses on protecting the machine and human from accidents and preventing the propagation of the risk from an accident during the operating campaign.

## II. MOTICATION FOR AN UPGRADE

The SIS is constituted by two horizontal layers, one for the central safety and interlock system (CSIS), and another for the different plant safety and interlock systems (PSIS), they are connected through safety and interlock network (SIN). All the systems in SIS are implemented by using SIEMENS PLC, they provide digital I/O channels with 1ms scan time, and the response time of interlock event is around 4ms. The SIN only consists of a set of optical fibres to transfer logic signals between CSIS and PSIS. With the development of physical experiment, the SIS had come close to reaching its limits for expandability. For instance, the former central safety and interlock system can't offer inspection channels less than 1ms scanning cycle. What more, the primitive GREEN and RED circles dashboard, and intermittent spurious monitor problem make the central control team determines to update the safety and interlock system.

The future needs of the EAST SIS research program are: (a) more stable supervisory and control operation; (b) high-precision execute and respond actions; (c) support multiple kinds of inspection objects; (d) better user-friendly interface [2]. Fulfilling all of these needs will require certain improvements to the previous capabilities including: (1) redesign the structure of SIS, (2) introduction of new equipment, (3) further development of the HMI, and (4) enhancements to the redundancy mechanism capability.

## III. SYSTEM STRUCTURE

The new SIS keeps the former two horizontal layers, and redistributes three vertical architectures according to timing requirements. The structure of the prototype SIS is illustrated in detail in Fig.1.

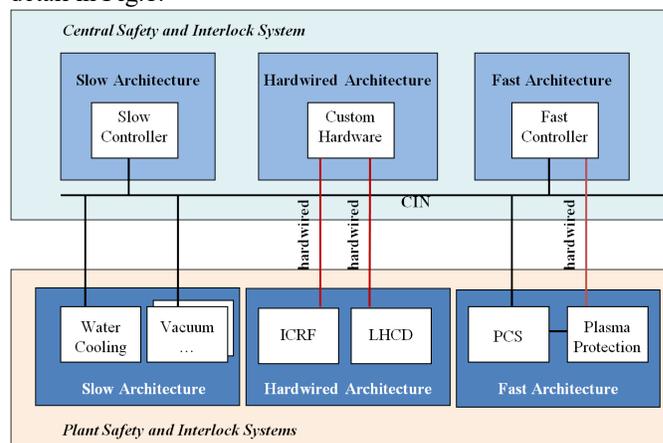

Fig. 1. Structure of the EAST SIS.

Each PSIS provides local protection implementing the local interlock functions of the corresponding plant system. The CIS is in charge of implementing the central protection functions via the PSIS through the central interlock network (CIN) and hardwired loop. Differing from SIN, the CIN not only provides connection between the CSIS and PSIS for inter-plant systems investment protection functions by optic fiber, but also provides access to the local interlock data of the different plant

Manuscript received October 30, 2020.
This work is supported by the Chinese National Key R&D Program of China No.2018YFE0302104 and No.2017YFE0300504. Z.C. Zhang, B.J. Xiao, Z.S. Ji, Y. Wang, Z.H. Xu are with the Department of Computer Application, Institute of Plasma Physics, Chinese Academy of Sciences, Hefei, Anhui, 230031, PR China (e-mail: zzc@ipp.ac.cn).
F. Xia is with the Southwestern Institute of Physics, Chengdu, Sichuan, 230031, PR China (e-mail: xiafan@swip.ac.cn).

interlock systems in the form of Ethernet. Communication within one plant system is carried out through the plant interlock network (PIN) and/or hardwired interconnections.

The systems such as water cooling, vacuum, etc. have slow response time requirements are categorized into slow architecture. System statuses are inspected through digital channels. For the functions of plasma safe stop and/or Poloidal Field (PF) coil current discharge in case an interlock event arises, plasma protection module has to be implemented on a fast architecture. Not only digital signals (interlock events and actions) but also analogue variables (plasma current, electron density etc.) are acquired. The third architecture is just used at Ion cyclotron range of frequency (ICRF) and lower hybrid current drive (LHCD). Threshold values are preset in custom hardware, when the acquired signals higher than values, the hardware will trigger the event generator and actuator

## IV. SYSTEM PROTECTION MECHANISM

The interface between the CSIS and plant systems is illustrated in Fig.2.

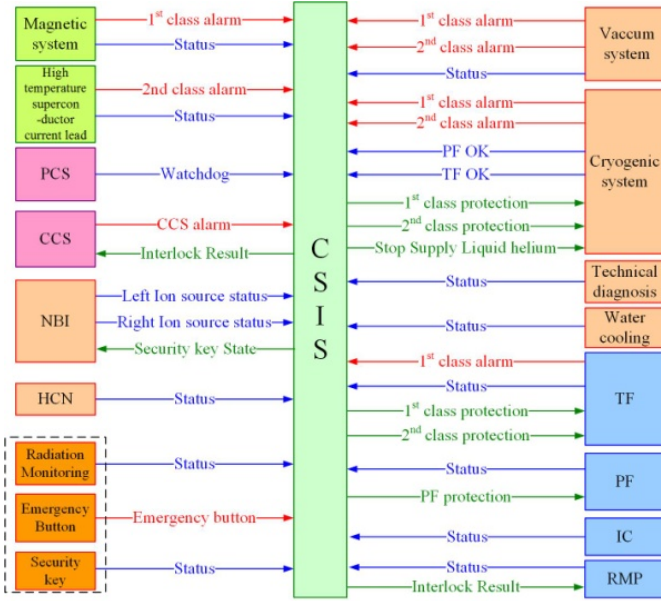

Fig. 2. Interface between CSIS and plant systems.

All interlock events managed by the CSIS are classified depending on their impact on the running or next plasma pulse. The relationship between event and response are shown in Table I.

TABLE I
RELATIONSHIP BETWEEN EVENT AND RESPONSE

| Signal | Value | Response |
|---|---|---|
| Status | High | OK |
|  | Low | Alarm |
| 1st Class alarm | High | 1st Class protection |
|  | Low | OK |
| 2nd Class alarm | High | 2nd Class protection |
|  | Low | OK |
| Protection | High | Actuator |
|  | Low | null |

## V. DESIGN OF SIS COMPONENTS

As mentioned, the SIS is divided in three parts according to timing requirements. The slow architecture has been established slower than 1ms based on PLC, and the fast one executes functions within 50μs by using NI COTS. The hardwired architecture has the shortest response time.

### A. Slow Architecture

The former SIS belongs to the slow architecture, and the system is intended to execute those central interlock functions with slow timing requirements, which have been established slower than 1ms. To increase the availability and reliability of the CSIS slow architecture, a redundant configuration has been selected. Fig.3 illustrated the structure of the CSIS slow architecture.

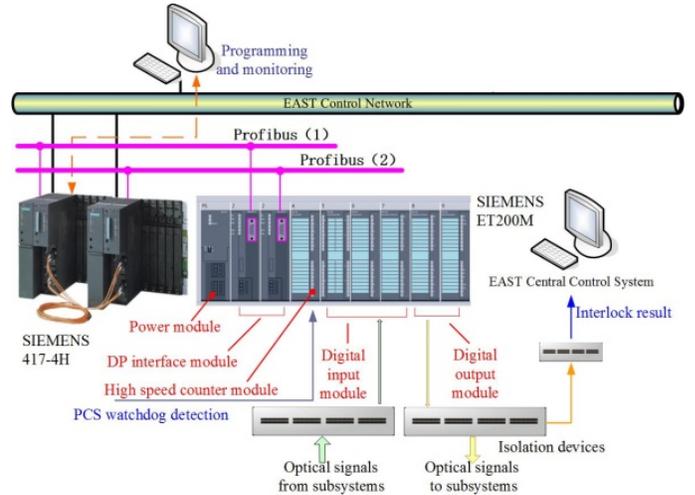

Fig. 3. Structure of the CSIS slow architecture.

SIEMENS SIMATIC S7-400 equipped with the following components:
• Master station:
   CPU 417-4H with 2 × 15MB Memory
   PS407 DC 24V/10A power supply
   A CP 443-1 advanced communication processor
• Slave station: ET200M
   PS307 DC 24V/10A power supply
   Two IM 153-2 Profibus-DP communication interfaces
   FM350-1 high-speed counter module
   Three SM 321 digital inputs DI 16 channels DC 24V
   Two SM 322 digital outputs DO 16 channels DC 24V.

All the signals from subsystems to the CSIS room are transmitted through fiber-optics. The isolation devices are custom devices which are used to remove ground loops among different subsystems. In Fig.2 it can be seen that the CSIS just monitors the status of subsystem with slow circle, so the EPICS is chosen to be deployed on the slow architecture. It should be specially explained that the slow architecture supervises the watch-dog signal from PCS by using high speed counter module, and the result only illustrates the status of PCS. The communication between PCS and protect module is described in fast architecture.

*B. Fast Architecture*

The EAST plasma control system operation cycle is 100μs, so the CSIS fast architecture is intended to execute the central safety and interlock functions in 50μs. From Fig.1, plasma protection module is the unique module in described CSIS fast architecture. This module executes all fast functions, as well as all functions that interface with PCS, which require a fast interface. It acquires plasma current, power supply (PS) over-current protection signal, and electron density, *etc*. Plasma actuators shall be also stopped by CSIS in case of complete loss of PCS. In parallel with the activation of disruption mitigation system, CSIS will stop all external heating and fuelling systems while ramping down to zero the current in all superconducting coils except TF.

This fast architecture has to perform interlock functions with an integrity level of up to 3IL-3 according to the IEC-61508 standard. The required integrity level should be achieved through different techniques (redundancy, failsafe communications, *etc*.). The ability of the user to reconfigure the interlock functions makes FPGA technology a good option [3].

As a difference to the slow architecture modules, the fast plasma protection module processes acquired both digital signals and analogue variables, which require certain calculations (e.g. plasma current and fusion power). The proposed solution to meet all the fast architecture requirements relies on the NI CompactRIO technology provided by National Instruments. NI CompactRIO is a reconfigurable FPGA-based embedded system for control and data acquisition applications. This architecture includes hot swappable I/O modules (C series) and a reconfigurable hardware system based on a chassis (housing a XILINX FPGA) in which modules are inserted [4]. NI FPGA platform is comprised by: (i)cRIO-9039, a 8-slot chassis features a Kintex-7 FPGA; (ii)two NI-9239, 4-channel AI, 24-Bit, ±10 V, 50 kS/s/ch ; (iii) two NI-9401, 8-channel DIO, 100 ns, 5V/TTL.

CA Lab, interface between LabVIEW and EPICS, makes the achievement of data simple now, this toolkit can create, read and write EPICS variables easily. EPICS time stamp, status, severity and optional PV fields (properties) are bound into a resulting data cluster. Users can get the data from both slow and fast architecture.

The human machine interface (HMI) of safety and interlock system is implemented by a server installed in the EAST acquisition room and through a client to be installed in central control room.

*C. Hardwired Architectures*

It is all known that the hardware circuit is the most responsive approach. Hardwired architecture in EAST CSIS is a set of custom devices, which consist of several sensors and actuators, having the capability of reading interlock event and generating response. Threshold value can be set through the rheostat. When the value of plasma current is lower than the preset threshold, the actuators will turn off the ability of heating system such as ICRF, LHCD. The majority of the functions are automatic, and the response time is less than 1μs.

## VI. Personal Safety Measure in CSIS

For personal safety, the safety part of CSIS provides five protection measures that are safety key, face recognition switch, water door, emergency button and radiation monitor. With the permission of operator, the person who enters the machine hall can get the safety key with a blue entrance guard card. Only experiment relevant person can pass the face recognition switch. Then water door system identifies the card's information and moves the water door to the specified location.

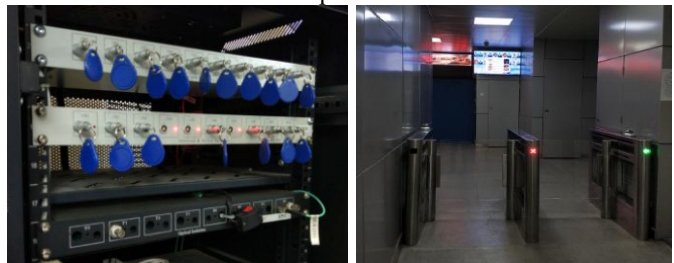

Fig. 4. Safety key and face recognition switch.

The EAST machine hall has two water doors. Two motors control these doors respectively [5]. Fig.5 is the framework of the water door's control system.

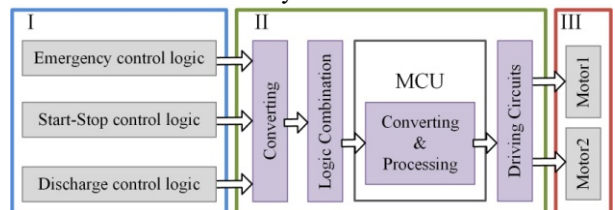

Fig. 5. Framework of water door's control system.

The control logic in Part I decides the states of motors' rotation and door's proper position. The Emergency control logic ensures that the gate can be opened quickly in an emergency case. Start-Stop logic enable the function of entrance guard cards identifies. Discharge control logic is a key control logic which comes from the EAST's discharge control signal. Part II is the control part which will output a proper control voltage to drive the motors based on the Part I's logic. The status of water door is also supervised in the CSIS slow architecture.

During the EAST discharge, some radioactive particles and rays will be emitted around the environment [6]. Fig.6 is the Neutron (n) and Gamma ray (c) detector.

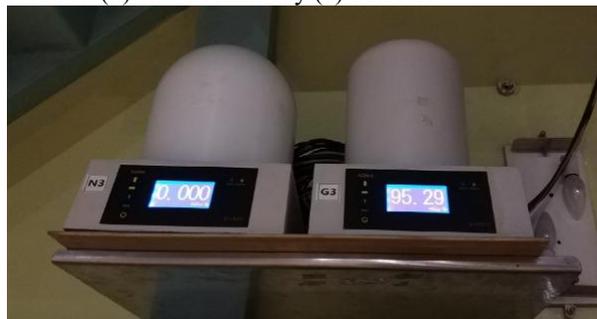

Fig. 6. Neutron (n) and Gamma ray (c) detector.

The probe of detector get the Neutron (n) and Gamma ray (c), outputs a voltage signal, after processing module, the result of radiation signal transmitted to the CSIS slow architecture.

## VII. Conclusion

At present, the EAST safety and interlock system works as described, with all the essential plant system included. SIEMENS PLC and National Instruments CompactRIO are integrated into a single system taking care of slow and fast interlock functions. The fast safety and interlock system is able to react within 50μs, and the personal safety modules provide reliable protection measures.

## Acknowledgment

The author would like to thank the EAST central control team and safety and interlock team for their work and help.

## References

[1] J. Xian, *et al.,* Nuclear radiation monitor based on single chip microcomputer, *Nucl. Electron. Detect. Technol.,* 1999(19).
[2] Z.S. Ji, Y.C. Wu, X. Sun, S. Li, F. Yang, Y. Wang, *et al.*, East integrated control system, *Fusion Eng. Des.,* vol. 85, pp. 509–514, May. 2011.
[3] T. Debelle and R. Marawar. (Mar. 2013). *Environmental Testing of NI Products for Big Physics Applications*. [Online]. Available: ftp://ftp.ni.com/pub/gdc/tut/environmental_test.pdf.
[4] E. Barrera, *et al.,* Implementation of ITER fast plant interlock system using FPGAs with CompactRIO, IEEE Transaction on Nuclear Science, vol. 65, no. 2, pp. 796–804, Feb. 2018.
[5] G.Z.Liu, L.Q.Hu, G. Q. Zhong, *et al.*, Control System of the EAST Hall's Gates, J. Fus. Eng ., vol. 34, no. 3,pp. 578–583, Jun. 2015.
[6] G.Z. Liu, L.Q. Hu, G. Q. Zhong, *et al.*, LabVIEW-Based Radiation Monitoring System of EAST, J. Fus. Eng ., vol. 35, no. 2,pp. 470–481, APR. 2016.